\pdfoutput=1
\pdfoutput=1
\pdfoutput=1
\pdfoutput=1
\pdfoutput=1

\documentclass[article]{achemso}

\usepackage[version=3]{mhchem} 
\usepackage[T1]{fontenc}       
\usepackage{caption}
\usepackage{float}
\usepackage{geometry}
\usepackage{natbib}
\usepackage{setspace}
\usepackage{xkeyval}
\usepackage{ifpdf}
\usepackage{graphicx}

\author{Sajjad AbdollahRamezani}
\affiliation{Department of Electrical Engineering, Sharif University of Technology, Tehran, Iran}
\author{Kamalodin Arik}
\affiliation{Department of Electrical Engineering, Sharif University of Technology, Tehran, Iran}
\author{Amin Khavasi}
\affiliation{Department of Electrical Engineering, Sharif University of Technology, Tehran, Iran}
\email{khavasi@sharif.edu}
\author{Zahra Kavehvash}
\affiliation{Department of Electrical Engineering, Sharif University of Technology, Tehran, Iran}

\title[An \textsf{achemso} demo]
{Analog Computing Using Graphene-based Metalines}


\begin{document}


\begin{abstract}
We introduce the new concept of "metalines" for manipulating the amplitude and phase profile of an incident wave locally and independently. Thanks to the highly confined graphene plasmons, a transmit-array of graphene-based metalines is used to realize analog computing on an ultra-compact, integrable and planar platform. By employing the general concepts of spatial Fourier transformation, a well-designed structure of such meta-transmit-array combined with graded index lenses can perform two mathematical operations; i.e. differentiation and integration, with high efficiency. The presented configuration is about $60$ times shorter than the recent structure proposed by Silva et al. (Science, 2014, 343, 160-163); moreover, our simulated output responses are in more agreement with the desired analytic results. These findings may lead to remarkable achievements in light-based plasmonic signal processors at nanoscale instead of their bulky conventional dielectric lens-based counterparts.
\end{abstract}



Recently, realization of analog computing has been achieved by manipulating continuous values of phase and amplitude of the transmitted and reflected waves by means of artificial engineered materials, known as metamaterials, and planar easy-to-fabricate metamaterials with periodic arrays of scaterrers, known as metasurfaces \cite{silva2014performing,pors2015analog,farmahini2013metasurfaces}. Both above-mentioned platforms offer the possibility of miniaturized wave-based computing systems that are several orders of magnitude thinner than conventional bulky lens-based optical processors \cite{silva2014performing,monticone2013full}.

Challenges associated with the complex fabrication of metamaterials \cite{engheta2006metamaterials,smith2004metamaterials,shalaev2007optical,monticone2014metamaterials,della2014digital} besides absorption loss of the metal constituent of metasurfaces \cite{kildishev2013planar,zhao2011manipulating,aieta2012aberration,saeidi2014wideband,saeidi2015figure}, degrade the quality of practical applications of relevant devices. As a result, graphene plasmonics can be a promising alternative due to the tunable conductivity of graphene and highly confined surface waves on graphene, the so-called graphene plasmons (GPs) \cite{vakil2011transformation,vakil2012fourier,tymchenko2013faraday,koppens2011graphene,bao2012graphene,lu2013flexible,chen2011atomically}.

\begin{figure}[t]
\centering
\includegraphics[trim=15.5cm 4.3cm 0cm 0cm,width=16.6cm,clip]{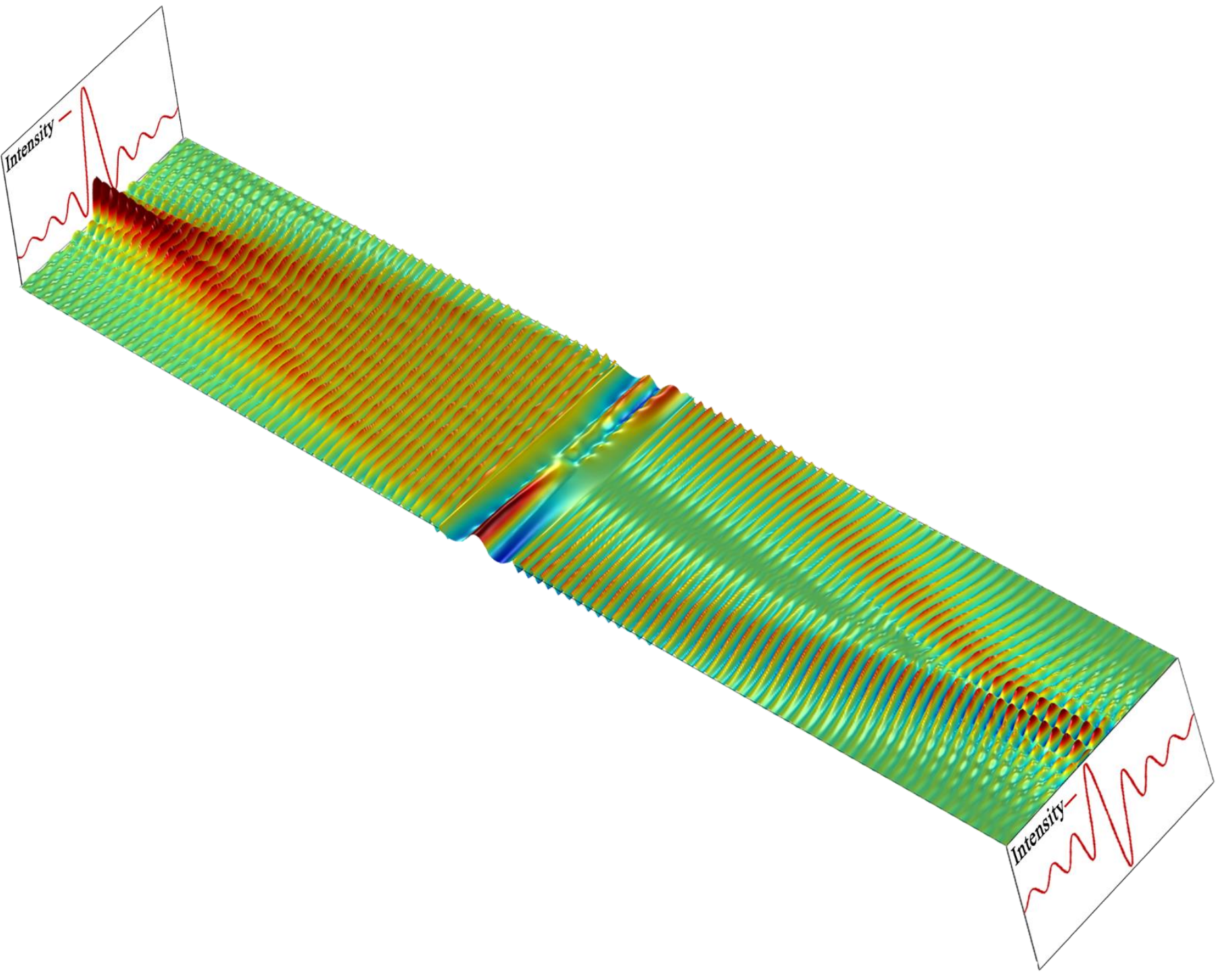}\\
\caption{Perspective view of wave propagation along a well-designed GRIN Lens/MTA/GRIN Lens which performs optical analog computing.}\label{fig0}
\end{figure}

We present a planar graphene-based configuration for manipulating GP waves to perform desired mathematical operations at nanoscale (see Figure~\ref{fig0}). By applying appropriate external gate voltage and a well-designed ground plane thickness profile beneath the dielectric spacer holding the graphene layer, desired surface conductivity values are achieved at different segments of graphene layer \cite{vakil2011transformation,fallahi2012design}. To illustrate the applications of the proposed configuration, two analog operators, i.e. differentiator and integrator, are designed and realized. The proposed structure will be two dimensional which is an advantage compared to the previously reported three dimensional structures \cite{silva2014performing,pors2015analog,farmahini2013metasurfaces} that manipulate one dimensional variable functions.

We introduce a new class of meta-transmit-arrays (MTA) on graphene: metalines which are one dimensional counterpart of metasurfaces. Our approach for realizing mathematical operators is similar to the first approach of [\cite{silva2014performing}]; i.e. metaline building blocks (instead of metasurfaces) perform mathematical operations in the spatial Fourier domain. However, the main advantage of our structure is that it is ultra-compact (its length is about $1/60$ of the free space wavelength, $\lambda$) in comparison with the structure proposed by Silva et al. whose length is about ${\lambda}/{3}$ \cite{silva2014performing}. It should be also noted that, in this work, the whole structure (including lenses and metalines) is implemented on graphene. Therefore, the total length of the proposed device is about $\lambda/4$, about $60$ times shorter than the device reported in [\cite{silva2014performing}].

\section*{THEORETICAL FRAMEWORK}

\begin{figure} 
\centering
\includegraphics[trim=0cm 8cm 0cm 0cm,width=16.6cm,clip]{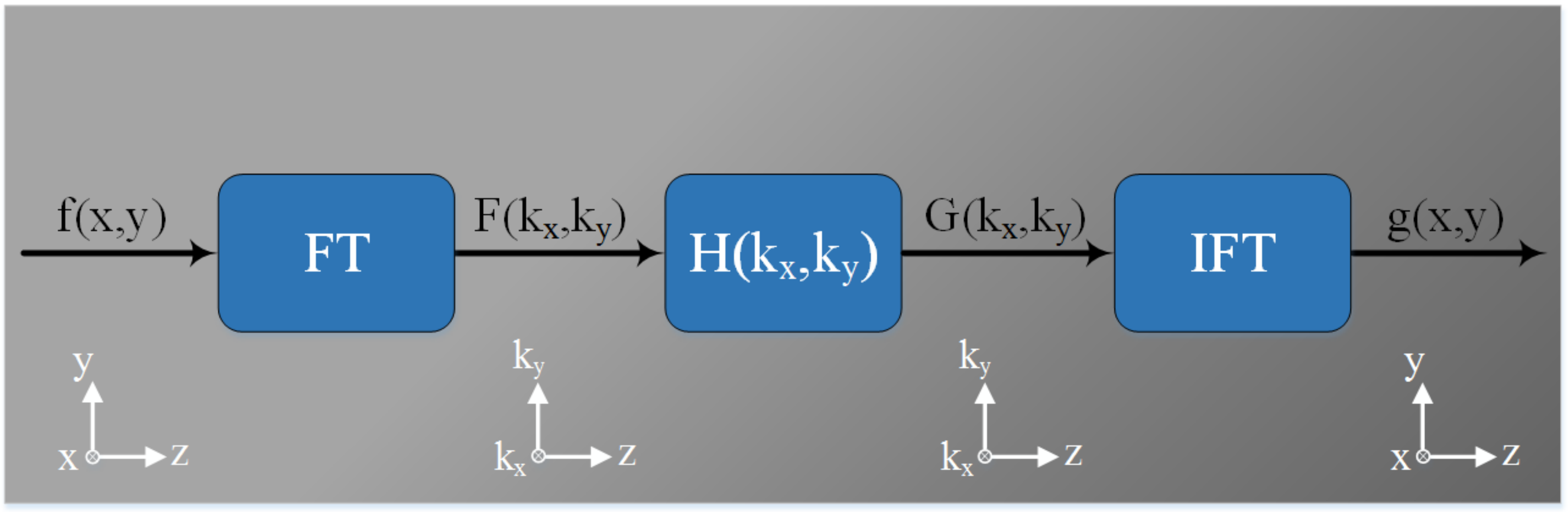}\\
\caption{Sketch of linear transversely invariant system to perform mathematical operations which consists of spatial Fourier transforming (FT), desired transfer function and inverse Fourier transforming (IFT) blocks.}\label{fig1}
\end{figure}

The general concept of performing mathematical operation in the spatial Fourier domain is graphically shown in Figure~\ref{fig1}. In this figure, $z$ is the propagation direction, $h(x,y)$ indicates the desired two dimensional impulse response, $f(x,y)$ is an arbitrary input function and $g(x,y)$ describes the corresponding output function. The whole system is assumed to be linear transversely invariant and thus the input and output functions are related to each other via the linear convolution \cite{pors2015analog}:
\begin{equation} 
g(x,y)=h(x,y){\ast}f(x,y)={\int}{\int}h(x-x',y-y')f(x',y')\textnormal{d}x'\textnormal{d}y'
\label{equ1}
\end{equation}
Transforming eq~\ref{equ1} to the spatial Fourier domain leads to:
\begin{equation} 
G(k_{x},k_{y})=H(k_{x},k_{y})F(k_{x},k_{y})
\label{equ2}
\end{equation}
where $G(k_{x},k_{y})$, $H(k_{x},k_{y})$ and $F(k_{x},k_{y})$ are the Fourier transform of their counterparts in eq~\ref{equ1}, respectively, and $(k_{x},k_{y})$ denotes the $2\textnormal{D}$ spatial Fourier domain variables.

For our two dimensional graphene-based system, as shown in Figure~\ref{fig2}, the above formulation is sufficient to be presented in one dimension. In this figure $f(x)$ and $g(x)$ represent transverse field distribution of the incident and transmitted waves, and $H(k_{x})$ is the appropriate transfer function. Accordingly, eq~\ref{equ2} can be interpreted as:
\begin{equation} 
g(x)=\textnormal{IFT}\lbrace{H(k_{x})\textnormal{FT}[f(x)]}\rbrace
\label{equ3}
\end{equation}
where $(\textnormal{I})\textnormal{FT}$ means (inverse) Fourier transform. The transfer function is given in the spatial Fourier domain $k_{x}$, and, on the other hand, the incident wave $(f(x))$ is also transformed into the Fourier domain. Hence, any transfer function can be realized by properly manipulating the Fourier-transformed wave in its transverse direction $x$ \cite{silva2014performing}.

Fourier transform can be carried out by a lens at its focal point. On the other hand, since realizing inverse Fourier transform with real materials is not possible, so instead of eq~\ref{equ3}, the following relation should be used:
\begin{equation} 
g(-x)=\textnormal{FT}\lbrace{H(k_{x})\textnormal{FT}[f(x)]}\rbrace
\label{equ4}
\end{equation}
This relation can be easily obtained from the well-known formula, $\textnormal{FT}\lbrace{\textnormal{FT}[g(x)]}\rbrace\propto{g(-x)}$. Eq~\ref{equ4} implies that the output will be proportional to the mirror image of the desired output function $g(x)$ \cite{silva2014performing}.

For performing Fourier transform we use graded index (GRIN) lenses. Since the optical properties of dielectric GRIN lenses change gradually, the scattering of the wave could be significantly reduced which leads to higher efficiency. To realize the graphene-based type of such lenses the surface conductivity of graphene should be properly patterned in a way that the effective mode index of GP waves follows the quadratic refractive index distribution of their dielectric counterparts \cite{wang2014graphene}.

\section*{GRAPHENE-BASED META-TRANSMITT-ARRAY}

In this work, we implement the appropriate transfer function, $H(k_{x})$ by means of a new type of meta-transmit-arrays on graphene. In order to manipulate the transmitted wave efficiently, not only the transmission amplitude should be completely controlled in the range of $0$ to $1$ \cite{pfeiffer2013cascaded,pfeiffer2014efficient}, but also transmission phase should cover the whole $2\pi$ range independently\cite{cheng2014optical,pors2013gap,pors2013plasmonic}. To this end, similar to the approach of Monticone et al. \cite{monticone2013full}, a meta-transmit-array comprised of symmetric stack of three metalines separated by a quarter-guided wavelength transmission line is utilized (Figure~\ref{fig2}c) in which each unit cell operates as a nanoscale spatial light modulator. To simplify the design procedure, the two outer stacks are chosen identical, but different from the inner one. The quarter-guided wavelength spacing between these stacks ensures that the transmission amplitude ripples are kept minimized \cite{lau2012reconfigurable}.

In order to fully control the transmission phase, in addition to amplitude, we need to locally manipulate propagating GP waves along and across the meta-transmit-array \cite{zentgraf2011plasmonic}. As described previously, this GP surface wave engineering is achieved via surface conductivity variation through an uneven ground plane beneath the graphene layer.

Recently, an analytical results for the reflection and transmission coefficients of GP waves at one dimensional surface conductivity discontinuity has been reported \cite{rejaei2015scattering}:
\begin{equation} 
r_{LR}=e^{i{\vartheta}_{LR}}\dfrac{k_{L}-k_{R}}{k_{L}+k_{R}},
t_{LR}=\dfrac{2(k_{L}k_{R})^{1/2}}{k_{L}+k_{R}}
\label{equ5}
\end{equation}
where
\begin{equation} 
{\vartheta}_{LR}=2\Psi(-k_{L})=\dfrac{\pi}{4}-\dfrac{2}{\pi}\int_{0}^{\infty}\dfrac{\textnormal{arctan}(k_{L}u/k_{R})}{u^{2}+1}\textnormal{d}u
\label{equ6}
\end{equation}
In these equations:
\begin{equation} 
k_{L,R}=\dfrac{2i\omega\varepsilon_{e}}{\sigma_{L,R}}
\label{equ7}
\end{equation}
are the GP wavenumbers of left and right side regions of the discontinuity in the quasi-static approximation, while $\sigma_{L,R}$ represent their corresponding complex surface conductivities, and $\varepsilon_{e}$ is the average permittivity of the upper and lower mediums surrounding the graphene sheet. Similarly, for a GP wave incident on the discontinuity from the right side region, these coefficients, $r_{RL}$ ans $t_{RL}$, can be simply achieved by exchanging $k_{L}$ and  $k_{R}$ in the  eqs~\ref{equ5},~\ref{equ6}. 

To relate the forward-backward fields on one side of the interface to those on the other side, the matching matrix is applied. 
By employing reciprocity theorem, the matching matrix of the interface is achieved as follows\cite{orfanidis2002electromagnetic}:
\begin{align}
M_{m}=
\begin{pmatrix} 
r_{LR,m}&t_{RL,m}\\
t_{LR,m}&r_{RL,m}
\end{pmatrix}
\label{equ8}
\end{align}
where m=0,$\ldots$,5 is the $m^{th}$ interface of the building block (see Figure~\ref{fig2}b). Furthermore, the propagative forward-backward fields relation along a segment are defined by propagation matrix \cite{orfanidis2002electromagnetic}:
\begin{align}
P_{m}=
\begin{pmatrix} 
e^{-ik_{m}l_{m}}&0\\
0&e^{ik_{m}l_{m}}
\end{pmatrix}
\label{equ9}
\end{align}
where m=1,$\ldots$,5 is the $m^{th}$ segment of the unit cell. Finally, the scattering parameters can be easily calculated using the whole building block transfer matrix obtained by multiplication of the matching and propagation matrices \cite{orfanidis2002electromagnetic}:
\begin{equation} 
T=M_{0}\prod_{m=1}^{5}P_{m}M_{m}
\label{equ10}
\end{equation}
The amplitude and phase of $S_{21}$ versus the chemical potentials of the inner and outer metalines, $\mu_{c,in}$ and $\mu_{c,out}$, are plotted in Figure~\ref{fig3}, calculated by means of our analytical approach. It is obvious that by local tuning of the chemical potential of each unit cell any transmission phase and amplitude profile can be achieved.

\section*{IMPLEMENTAION OF MATHEMATICAL OPERATORS AND REPRESENTATIVE RESULTS }

In this section, we implement first differentiation, second differentiation and integration operators, with the proposed structure. It is well-known that the $n^{th}$ derivative of a function is related to its first Fourier transform by:
\begin{equation} 
\dfrac{\textnormal{d}^{n}(f(x))}{\textnormal{d}x^{n}}=\textnormal{F}^{-1}\lbrace(ik_{x})^{n}\textnormal{F}(f(x))\rbrace
\label{equ11}
\end{equation}
Comparing eq~\ref{equ3} and eq~\ref{equ11} clarifies that to realize the $n^{th}$ derivation we have to perform a transfer function of $(ik_{x})^{n}$. So as described in the previous section we set the transfer function of the meta-transmit-array to $H(x)\propto(ix)^{n}$. Since metalines are inherently passive media, the desired transfer function has to be normalized to the lateral limit to ensure that across the structure the maximum transmittance is unity, thus the appropriate transfer function is $H(x)\propto(ix/(W/2))^{n}$. Figures~\ref{fig4}b,c indicate the desired magnitude and phase profile of the first-order derivative transfer function $H(x)$. According to Figure~\ref{fig3}, by properly tailoring the values of the chemical potential along the lateral dimension for internal and external metalines, the transverse amplitude and phase distribution of the transfer function are implemented. 
To this end, the transverse distribution of chemical potential for the designed first-order differentiator meta-transmit-array and its corresponding complex surface conductivity profile are depicted in Figure \ref{fig10}.
Now, a TM-polarized GP surface wave is launched toward the designed meta-transmit-array and the calculated electric field distribution is shown in Figure~\ref{fig4}. As depicted in Figures~\ref{fig4}b,c the output profile of meta-transmit-array is in an excellent agreement with the desired transfer function.
To be more precise, the standard deviation from the amplitude and phase of the desired transfer function is $0.04$ and $6^{\circ}$, respectively. This is smaller than the standard deviation observed in Fig. S7b of [\cite{silva2014performing}] which is about $0.19$ and $15^{\circ}$ for the amplitude and phase, respectively. As a result, a more accurate output response can be anticipated for the whole structures.

Other operators can be designed in a similar manner. In order to perform the second-order spatial derivative, the desired transfer function is $H(x)\propto-(x/(W/2))^{2}$. Although here the amplitude is a quadratic function of transverse dimension, the phase is constant. The results for the designed second-order differentiator are illustrated in Figure~\ref{fig5}.

To realize a second-order integrator, a challenge should be overcome. In this case, the desirable transfer function should be $H(x)\propto(ix)^{-2}$ which leads to an amplitude profile with values tending to infinity in the vicinity of $x=0$. To overcome this problem, we use the following approximate transfer function \cite{silva2014performing}:
\begin{equation}
  H(x)=\begin{cases}
    \hspace{3mm}1\hspace{6mm}, & \text{if $\vert{x}\vert<h$}\\
    (ix)^{-2}, & \text{if $\vert{x}\vert>h$}
  \end{cases}
  \label{equ12}
\end{equation}
where $h$ is an arbitrary parameter that we set it to $h=W/12$ \cite{silva2014performing}. For all the points within $\vert{x}\vert<h$  the amplitude is assumed to be unity, others follow the correct transfer function profile precisely. Figure~\ref{fig6}a shows numerical simulation of electric field distribution for this case. By comparing the obtained results from meta-transmit-array and analytical solution in Figures~\ref{fig6}b,c , an excellent agreement is observed.
 
To demonstrate the functionality of our proposed structures, a GP surface wave in the form of a Sinc function is sourced into our proposed GRIN/meta-transmit-array/GRIN configuration and the simulated electric field distributions are illustrated in figures~\ref{fig7}, \ref{fig8} and \ref{fig9} for the designed first-order derivative, second-order derivative and second order integrator, respectively. It is obvious from these figures that the achieved results are closely proportional to the desired results calculated analytically.

\section*{CONCLUSIONS AND OUTLOOK}

In summary, we have proposed and designed a new class of planar meta-transmit-array consisting of symmetric three stacked graphene-based metalines to perform wave-based analog computing. Using analytical results for the reflection and transmission coefficients of graphene plasmon waves at one dimensional surface conductivity discontinuity \cite{rejaei2015scattering}, we have demonstrated that full control over the transmission amplitude and phase can be achieved by appropriately tailoring of surface conductivity of each building block. Employing general concept of performing mathematical operations in spatial Fourier domain, we assign the meta-transmit-array a specific transfer function corresponding to the desired operation. The two designed operators in this work; i.e. differentiator and integrator, illustrate a high efficiency. 
The proposed graphene-based structure not only is ultra-compact, but also depicts more accurate responses than the bulky structure suggested in [\cite{silva2014performing}]. These features are due to exceptionally high confinement of surface plasmons propagating on a graphene sheet. 
However, this miniature size comes at a price, namely any future fabrication imperfections and tolerance will lead to distortion of the anticipated results, and therefore, degrade the efficiency of the proposed structure.
The presented approach may broaden horizons to achieve more complex nanoscale signal processors.

\section*{MATERIALS AND METHODS}

The full-wave simulations are carried out by the commercial electromagnetic solver, Ansoft's HFSS. Since a high confined GP wave is stimulated here, in all our numerical simulations a free standing graphene layer is assumed \cite{vakil2011transformation}. In the simulations, the graphene layer is treated as an surface impedance $Z_{s}=1/\sigma_{s}$, where $\sigma_{s}$ is the complex surface conductivity of graphene can be retrieved by the Kubo's formula under random phase approximation \cite{hadad2014extreme}:
\begin{align} 
\sigma_{s}&=\dfrac{ie^{2}k_{B}T}{\pi\hbar^{2}(\omega+i2\Gamma)}[\dfrac{\mu_{c}}{k_{B}T}+2\ln(e^{(-\mu_{c}/k_{B}T)}+1)] \nonumber \\
&+\dfrac{ie^{2}}{4\pi\hbar}\ln[\dfrac{2|\mu_{c}|-(\omega+i2\Gamma)\hbar}{2|\mu_{c}|+(\omega+i2\Gamma)\hbar}]
\label{equ13}
\end{align}
where $\omega$ is the angular frequency, $\mu_{c}$ is the chemical potential, $\tau=1/2{\Gamma}$ is the relaxation time which represents loss mechanism, $T$ is the temperature, $e$ is the charge of an electron, $\hbar$ is the reduced Planck's constant, and $k_{B}$ is Boltzmann's constant. In our work, $T=300$ K and $\tau=1$ ps, which are relevant to the recently experimental achievements\cite{shen2015tunable}. Moreover, A time dependence of the form $e^{-i{\omega}t}$ is assumed and suppressed throughout this study.

Due to very small mesh size, the simulation becomes computationally time consuming. As a result, we follow the approach in \cite{farmahini2013metasurfaces}, i.e. block-by-block simulation. Initially, the first GRIN lens is excited with the desired input field profile. Then, the meta-transmit-array is excited by the monitored output field distribution of the first block. At last, the obtained output of the MTA is sourced into the second GRIN lens. To excite the GP surface wave with the desired field distribution, a current sheet is placed perpendicular to the graphene layer. By discretizing the current sheet to sufficient number of segments and assigning desired surface current amplitude and phase to them, each segment acts as an independent source. Consequently, any exact field distribution can be achieved easily.

\begin{figure} 
\centering
\includegraphics[trim=3cm 4.4cm 0cm 0cm,width=16.6cm,clip]{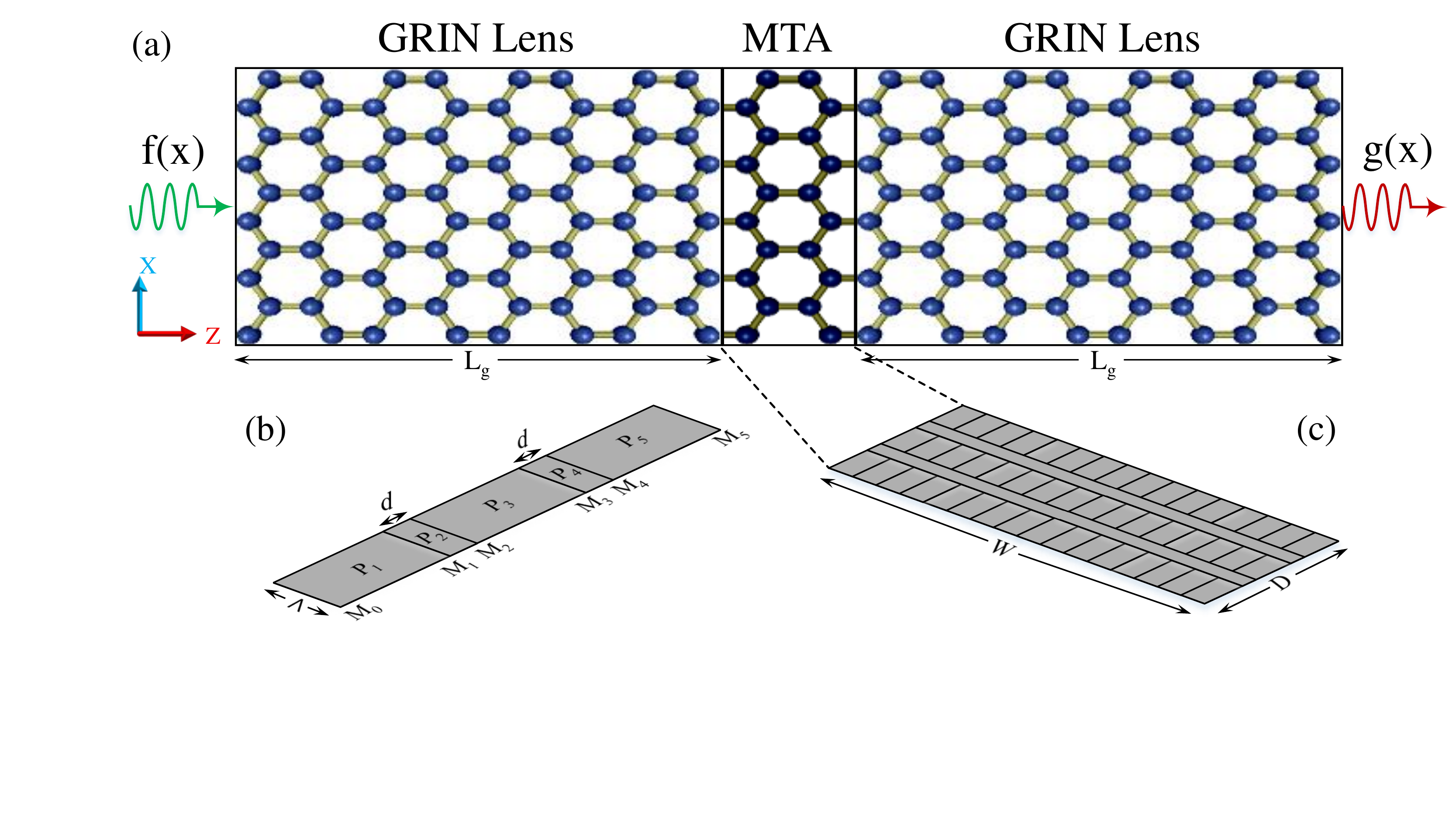}\\
\caption{(a) Schematic of two dimensional graphene-based computing system including GRIN Lens/MTA/GRIN Lens together with the input and output functions which variate across $x$ direction and propagate along $z$ direction. (b) Basic building block of the metalines labeled by the propagation and matching matrices corresponding to the interfaces and segments. (c) Sketch of the meta-transmit-array made of three symmetric stacked metalines embedded between two GRIN Lenses.}\label{fig2}
\end{figure}

\begin{figure}
\centering
\includegraphics[trim=11.5cm 9.5cm 0cm 0cm,width=16.6cm,clip]{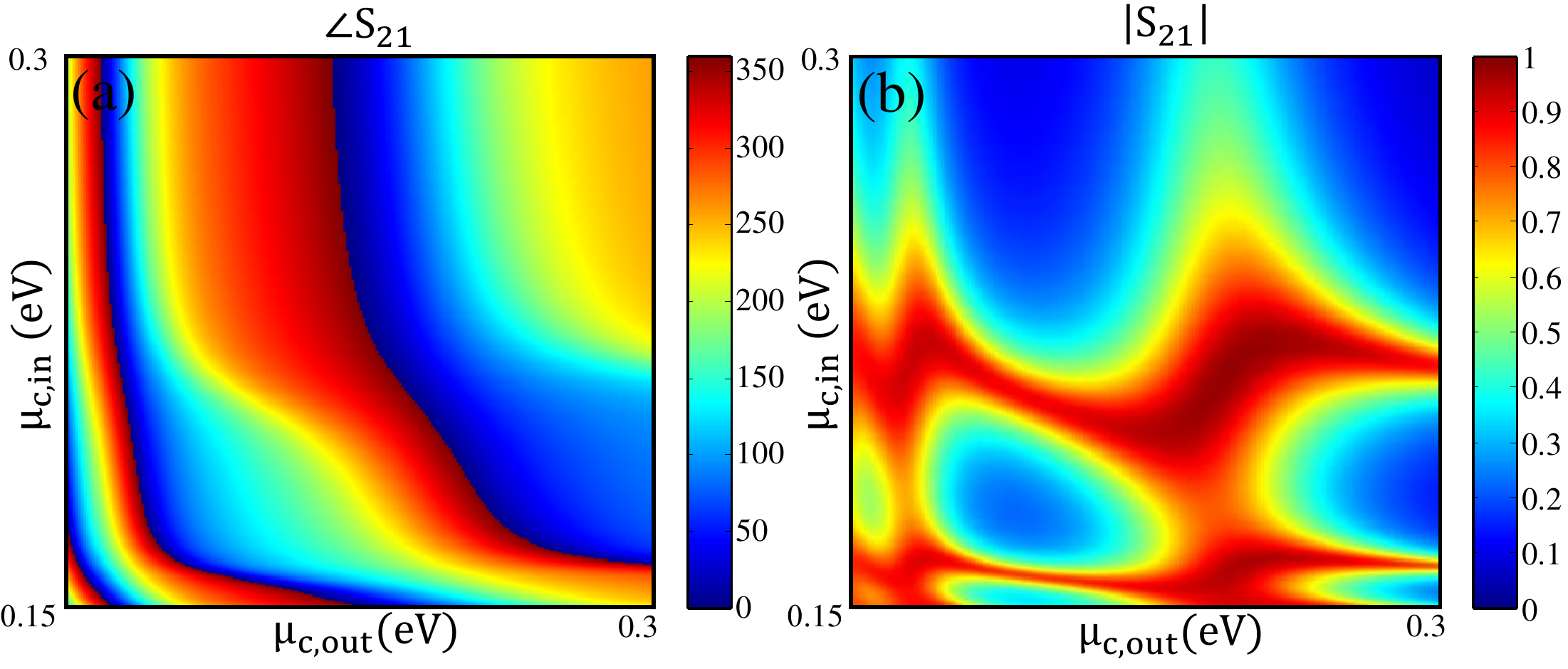}\\
\caption{(a) Phase and (b) amplitude of transmission coefficient versus the internal and external metalines' chemical potentials calculated by the proposed analytical approach.}\label{fig3}
\end{figure}

\begin{figure} 
\centering
\includegraphics[trim=9cm 0cm 9cm 0cm,width=16.6cm,clip]{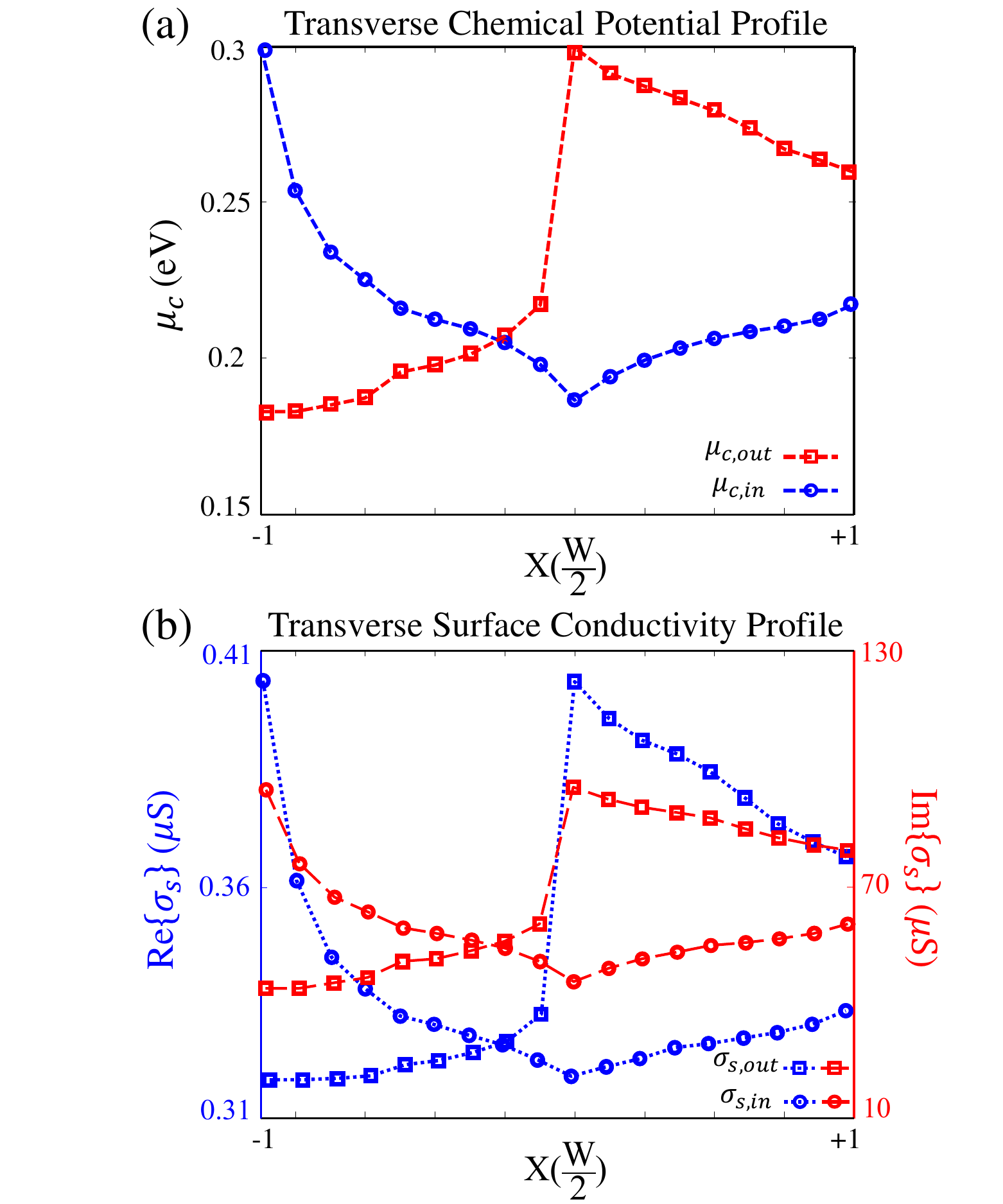}\\
\caption{The (a) transverse distribution of chemical potential for the designed first-order differentiator meta-transmit-array and (b) corresponding complex surface conductivity.}\label{fig10}
\end{figure}

\begin{figure} 
\centering
\includegraphics[trim=8.6cm 0cm 0cm 0cm,width=16.6cm,clip]{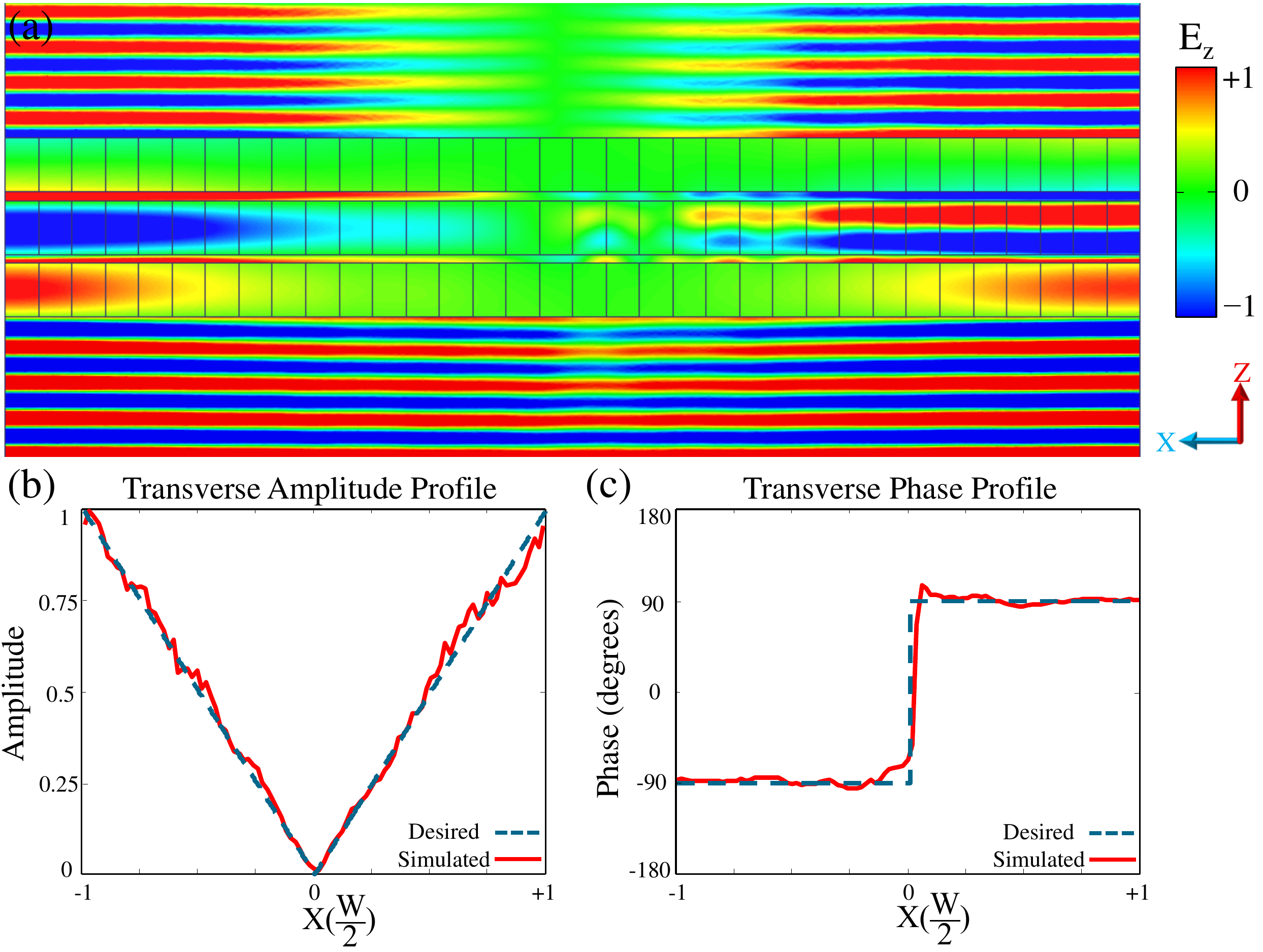}\\
\caption{(a) Snapshot of electric field distribution $(E_{z})$ for a TM-polarized GP surface wave incident on the designed first-order differentiator meta-transmit-array. Comparison of the transverse (b) amplitude and (c) phase distribution of transmitted wave and desired transfer function response right behind the structure.}\label{fig4}
\end{figure}

\begin{figure} 
\centering
\includegraphics[trim=8.6cm 0cm 0cm 0cm,width=16.6cm,clip]{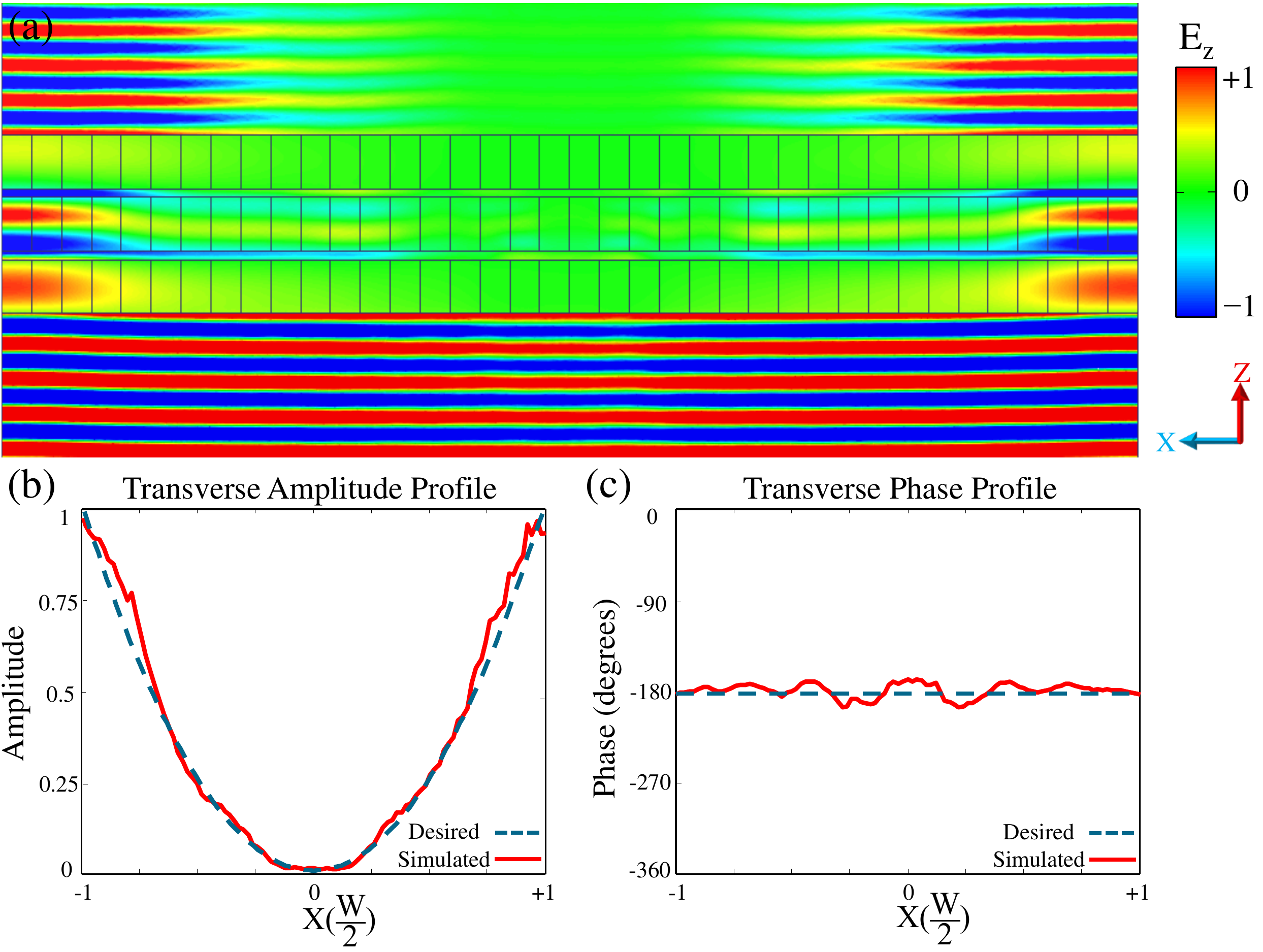}\\
\caption{(a) Snapshot of electric field distribution $(E_{z})$ for a TM-polarized GP surface wave incident on the designed second-order differentiator meta-transmit-array. Comparison of the transverse (b) amplitude and (c) phase distribution of transmitted wave and desired transfer function response right behind the structure.}\label{fig5}
\end{figure}

\begin{figure} 
\centering
\includegraphics[trim=8.6cm 0cm 0cm 0cm,width=16.6cm,clip]{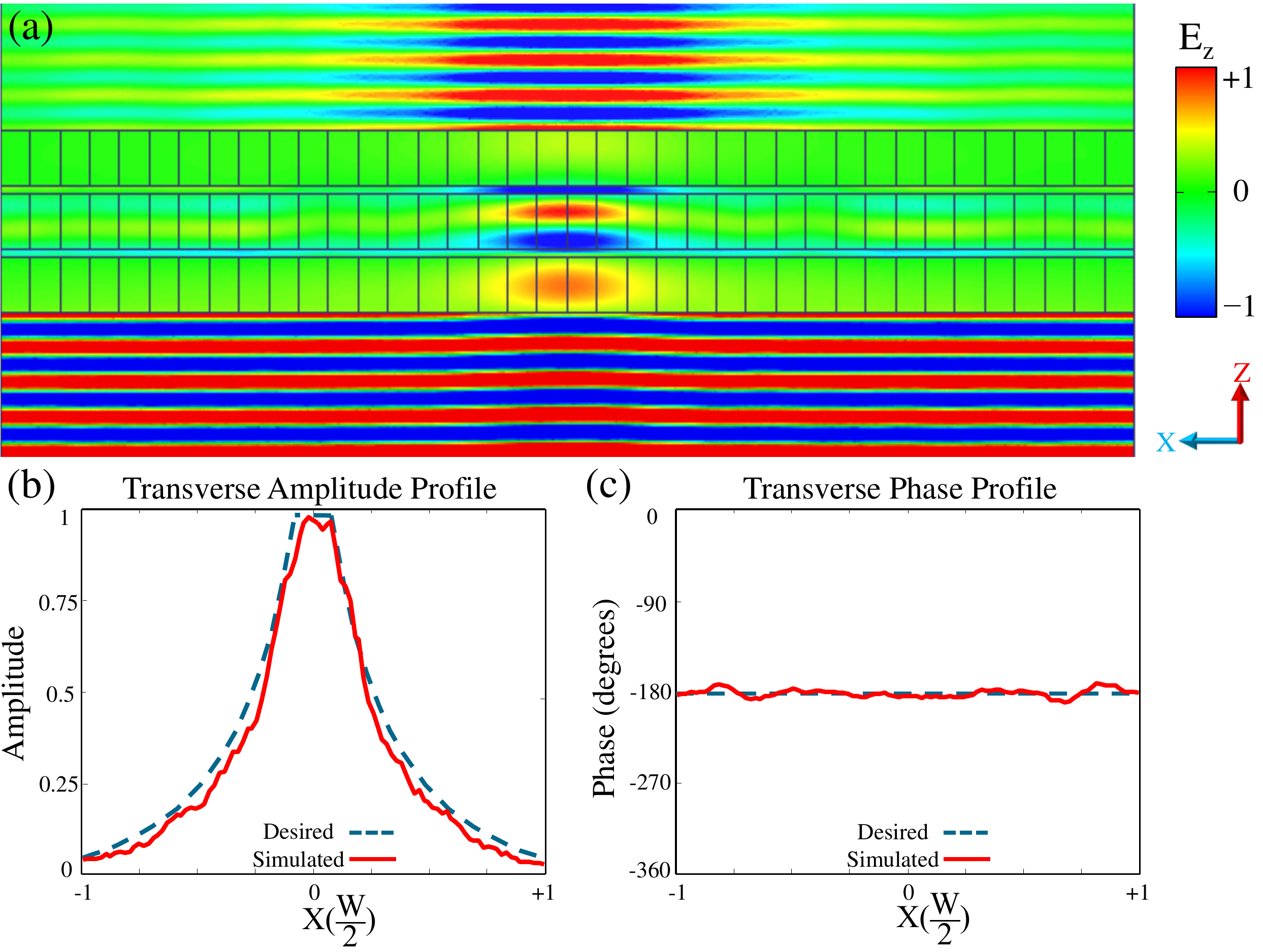}\\
\caption{(a) Snapshot of electric field distribution $(E_{z})$ for a TM-polarized GP surface wave incident on the designed second-order integrator meta-transmit-array. Comparison of the transverse (b) amplitude and (c) phase distribution of transmitted wave and desired transfer function response right behind the structure.}\label{fig6}
\end{figure}

\begin{figure} 
\centering
\includegraphics[trim=14.7cm 0cm 0cm 0cm,width=16.6cm,clip]{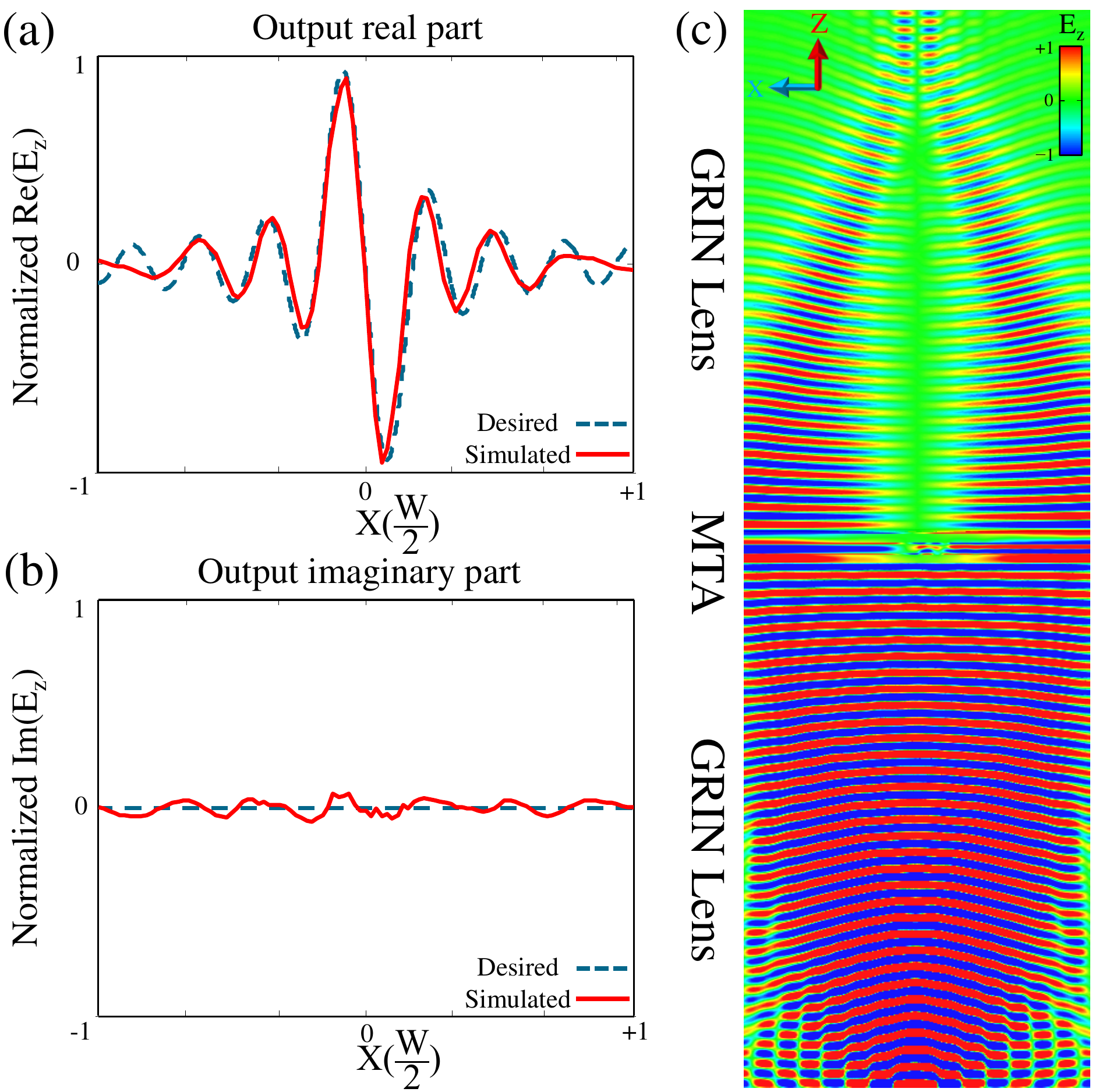}\\
\caption{The (a) real and (b) imaginary parts of the output electric field compared with the analytical results for the first-order differentiator. (c) Snapshot of the $z$-component of the electric field distribution along the GRIN Lens/MTA/GRIN Lens. The input function is $f(x)=\textnormal{sinc}(16{\pi}x/W)$.}\label{fig7}
\end{figure}

\begin{figure} 
\centering
\includegraphics[trim=14.7cm 0cm 0cm 0cm,width=16.6cm,clip]{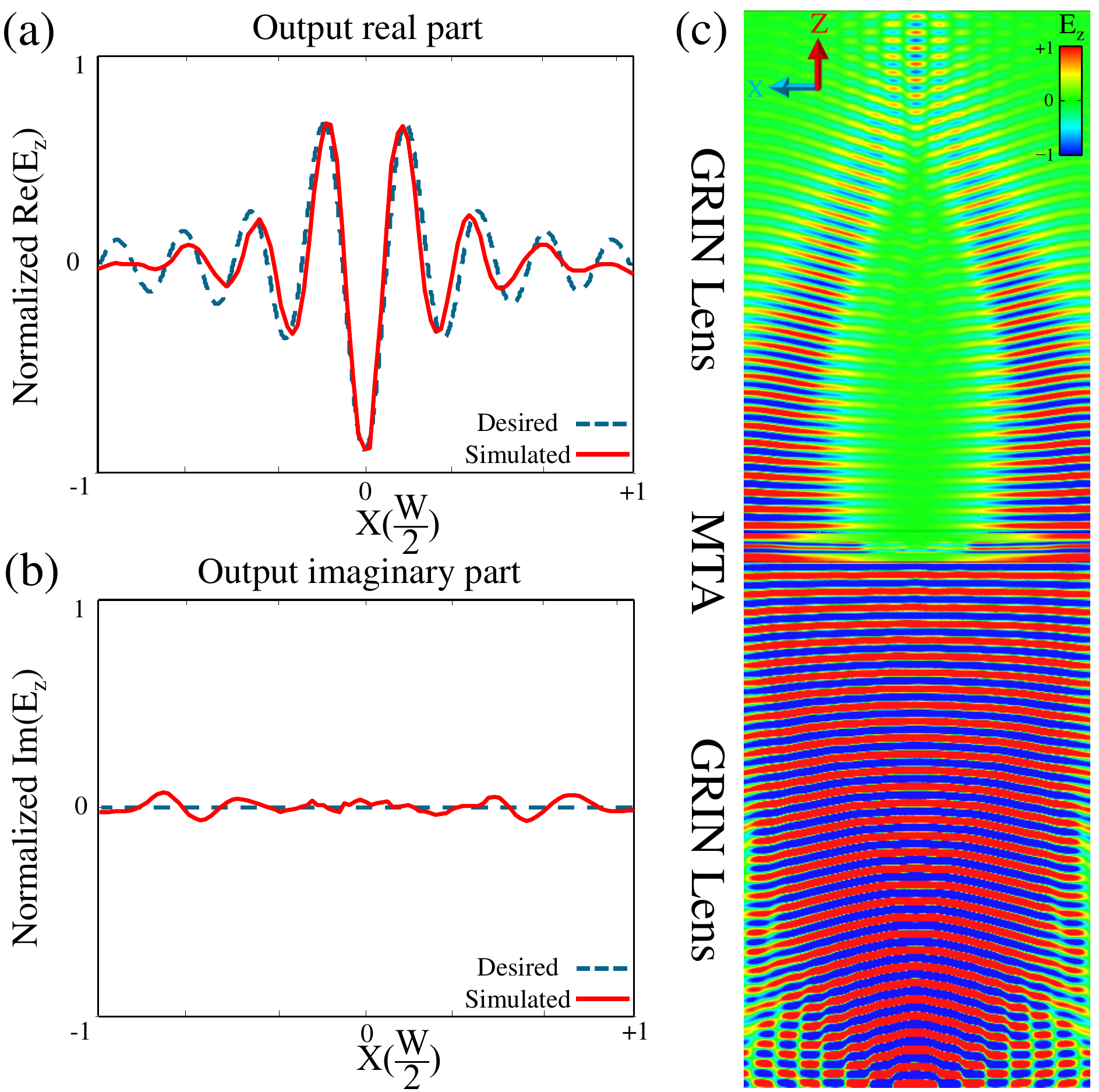}\\
\caption{The (a) real and (b) imaginary parts of the output electric field compared with the analytical results for the second-order differentiator. (c) Snapshot of the $z$-component of the electric field distribution along the GRIN Lens/MTA/GRIN Lens. The input function is $f(x)=\textnormal{sinc}(16{\pi}x/W)$.}\label{fig8}
\end{figure}

\begin{figure} 
\centering
\includegraphics[trim=14.7cm 0cm 0cm 0cm,width=16.6cm,clip]{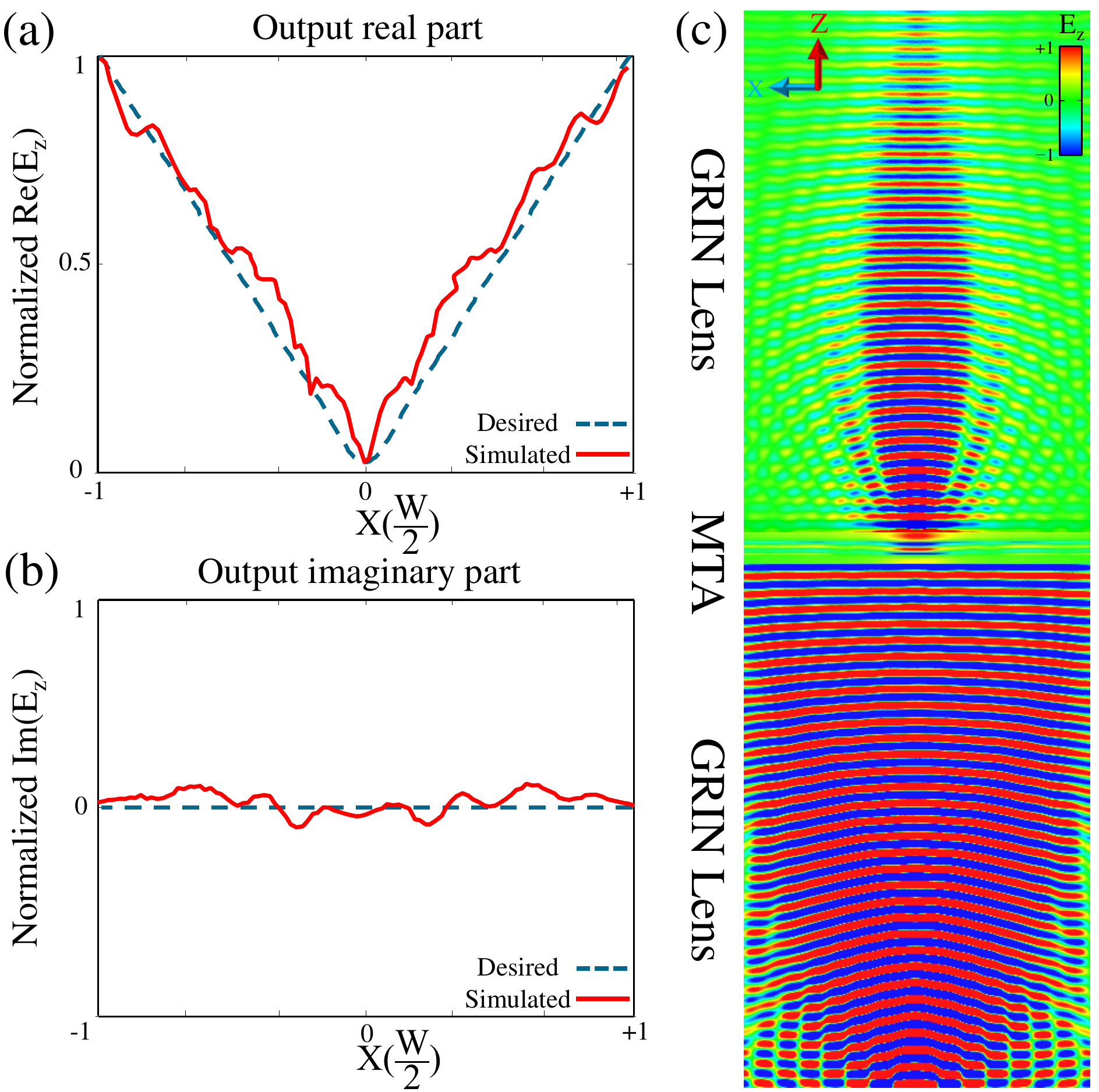}\\
\caption{The (a) real and (b) imaginary parts of the output electric field compared with the analytical results for the second-order integrator. (c) Snapshot of the $z$-component of the electric field distribution along the GRIN Lens/MTA/GRIN Lens. The input function is $f(x)=\textnormal{sinc}(16{\pi}x/W)$.}\label{fig9}
\end{figure}

The schematic model used in the simulations is shown in Figure~\ref{fig2} which contains two GRIN lenses and the meta-transmit-array embedded between them. The whole structure is restricted in transversal and longitudinal directions by $W$ and $2L_{g}+D$, respectively. Here, we set $W=684$ nm, $L_{g}=1028$ nm, $D=100$ nm, $\Lambda=18$ nm, $d=5$ nm. In all simulated examples, the wavelength is chosen $\lambda=6$ $\mu$m. The proposed configuration can principally be implemented for other wavelengths using well-designed constituent parameters and dimensions, as long as the graphene layer supports the GP waves at that wavelength. A $y$-polarized GP surface wave with spatial variation $f(x)=\textnormal{sinc}(16{\pi}x/W)$ is used as the input function to evaluate the efficiency of the whole GRIN Lens/MTA/GRIN Lens structure.
\ref{fig10}


\providecommand{\latin}[1]{#1}
\providecommand*\mcitethebibliography{\thebibliography}
\csname @ifundefined\endcsname{endmcitethebibliography}
  {\let\endmcitethebibliography\endthebibliography}{}
\begin{mcitethebibliography}{35}
\providecommand*\natexlab[1]{#1}
\providecommand*\mciteSetBstSublistMode[1]{}
\providecommand*\mciteSetBstMaxWidthForm[2]{}
\providecommand*\mciteBstWouldAddEndPuncttrue
  {\def\EndOfBibitem{\unskip.}}
\providecommand*\mciteBstWouldAddEndPunctfalse
  {\let\EndOfBibitem\relax}
\providecommand*\mciteSetBstMidEndSepPunct[3]{}
\providecommand*\mciteSetBstSublistLabelBeginEnd[3]{}
\providecommand*\EndOfBibitem{}
\mciteSetBstSublistMode{f}
\mciteSetBstMaxWidthForm{subitem}{(\alph{mcitesubitemcount})}
\mciteSetBstSublistLabelBeginEnd
  {\mcitemaxwidthsubitemform\space}
  {\relax}
  {\relax}

\bibitem[Silva \latin{et~al.}(2014)Silva, Monticone, Castaldi, Galdi, Al{\`u},
  and Engheta]{silva2014performing}
Silva,~A.; Monticone,~F.; Castaldi,~G.; Galdi,~V.; Al{\`u},~A.; Engheta,~N.
  Performing mathematical operations with metamaterials. \emph{Science}
  \textbf{2014}, \emph{343}, 160--163\relax
\mciteBstWouldAddEndPuncttrue
\mciteSetBstMidEndSepPunct{\mcitedefaultmidpunct}
{\mcitedefaultendpunct}{\mcitedefaultseppunct}\relax
\EndOfBibitem
\bibitem[Pors \latin{et~al.}(2015)Pors, Nielsen, and
  Bozhevolnyi]{pors2015analog}
Pors,~A.; Nielsen,~M.~G.; Bozhevolnyi,~S.~I. Analog computing using reflective
  plasmonic metasurfaces. \emph{Nano letters} \textbf{2015}, \relax
\mciteBstWouldAddEndPunctfalse
\mciteSetBstMidEndSepPunct{\mcitedefaultmidpunct}
{}{\mcitedefaultseppunct}\relax
\EndOfBibitem
\bibitem[Farmahini-Farahani \latin{et~al.}(2013)Farmahini-Farahani, Cheng, and
  Mosallaei]{farmahini2013metasurfaces}
Farmahini-Farahani,~M.; Cheng,~J.; Mosallaei,~H. Metasurfaces nanoantennas for
  light processing. \emph{JOSA B} \textbf{2013}, \emph{30}, 2365--2370\relax
\mciteBstWouldAddEndPuncttrue
\mciteSetBstMidEndSepPunct{\mcitedefaultmidpunct}
{\mcitedefaultendpunct}{\mcitedefaultseppunct}\relax
\EndOfBibitem
\bibitem[Monticone \latin{et~al.}(2013)Monticone, Estakhri, and
  Al{\`u}]{monticone2013full}
Monticone,~F.; Estakhri,~N.~M.; Al{\`u},~A. Full control of nanoscale optical
  transmission with a composite metascreen. \emph{Physical Review Letters}
  \textbf{2013}, \emph{110}, 203903\relax
\mciteBstWouldAddEndPuncttrue
\mciteSetBstMidEndSepPunct{\mcitedefaultmidpunct}
{\mcitedefaultendpunct}{\mcitedefaultseppunct}\relax
\EndOfBibitem
\bibitem[Engheta and Ziolkowski(2006)Engheta, and
  Ziolkowski]{engheta2006metamaterials}
Engheta,~N.; Ziolkowski,~R.~W. \emph{Metamaterials: physics and engineering
  explorations}; John Wiley \& Sons, 2006\relax
\mciteBstWouldAddEndPuncttrue
\mciteSetBstMidEndSepPunct{\mcitedefaultmidpunct}
{\mcitedefaultendpunct}{\mcitedefaultseppunct}\relax
\EndOfBibitem
\bibitem[Smith \latin{et~al.}(2004)Smith, Pendry, and
  Wiltshire]{smith2004metamaterials}
Smith,~D.~R.; Pendry,~J.~B.; Wiltshire,~M.~C. Metamaterials and negative
  refractive index. \emph{Science} \textbf{2004}, \emph{305}, 788--792\relax
\mciteBstWouldAddEndPuncttrue
\mciteSetBstMidEndSepPunct{\mcitedefaultmidpunct}
{\mcitedefaultendpunct}{\mcitedefaultseppunct}\relax
\EndOfBibitem
\bibitem[Shalaev(2007)]{shalaev2007optical}
Shalaev,~V.~M. Optical negative-index metamaterials. \emph{Nature photonics}
  \textbf{2007}, \emph{1}, 41--48\relax
\mciteBstWouldAddEndPuncttrue
\mciteSetBstMidEndSepPunct{\mcitedefaultmidpunct}
{\mcitedefaultendpunct}{\mcitedefaultseppunct}\relax
\EndOfBibitem
\bibitem[Monticone and Alu(2014)Monticone, and Alu]{monticone2014metamaterials}
Monticone,~F.; Alu,~A. Metamaterials and plasmonics: From nanoparticles to
  nanoantenna arrays, metasurfaces, and metamaterials. \emph{Chin. Phys. B}
  \textbf{2014}, \emph{23}, 047809\relax
\mciteBstWouldAddEndPuncttrue
\mciteSetBstMidEndSepPunct{\mcitedefaultmidpunct}
{\mcitedefaultendpunct}{\mcitedefaultseppunct}\relax
\EndOfBibitem
\bibitem[Della~Giovampaola and Engheta(2014)Della~Giovampaola, and
  Engheta]{della2014digital}
Della~Giovampaola,~C.; Engheta,~N. Digital metamaterials. \emph{Nature
  materials} \textbf{2014}, \emph{13}, 1115--1121\relax
\mciteBstWouldAddEndPuncttrue
\mciteSetBstMidEndSepPunct{\mcitedefaultmidpunct}
{\mcitedefaultendpunct}{\mcitedefaultseppunct}\relax
\EndOfBibitem
\bibitem[Kildishev \latin{et~al.}(2013)Kildishev, Boltasseva, and
  Shalaev]{kildishev2013planar}
Kildishev,~A.~V.; Boltasseva,~A.; Shalaev,~V.~M. Planar photonics with
  metasurfaces. \emph{Science} \textbf{2013}, \emph{339}, 1232009\relax
\mciteBstWouldAddEndPuncttrue
\mciteSetBstMidEndSepPunct{\mcitedefaultmidpunct}
{\mcitedefaultendpunct}{\mcitedefaultseppunct}\relax
\EndOfBibitem
\bibitem[Zhao and Al{\`u}(2011)Zhao, and Al{\`u}]{zhao2011manipulating}
Zhao,~Y.; Al{\`u},~A. Manipulating light polarization with ultrathin plasmonic
  metasurfaces. \emph{Physical Review B} \textbf{2011}, \emph{84}, 205428\relax
\mciteBstWouldAddEndPuncttrue
\mciteSetBstMidEndSepPunct{\mcitedefaultmidpunct}
{\mcitedefaultendpunct}{\mcitedefaultseppunct}\relax
\EndOfBibitem
\bibitem[Aieta \latin{et~al.}(2012)Aieta, Genevet, Kats, Yu, Blanchard,
  Gaburro, and Capasso]{aieta2012aberration}
Aieta,~F.; Genevet,~P.; Kats,~M.~A.; Yu,~N.; Blanchard,~R.; Gaburro,~Z.;
  Capasso,~F. Aberration-free ultrathin flat lenses and axicons at telecom
  wavelengths based on plasmonic metasurfaces. \emph{Nano letters}
  \textbf{2012}, \emph{12}, 4932--4936\relax
\mciteBstWouldAddEndPuncttrue
\mciteSetBstMidEndSepPunct{\mcitedefaultmidpunct}
{\mcitedefaultendpunct}{\mcitedefaultseppunct}\relax
\EndOfBibitem
\bibitem[Saeidi and van~der Weide(2014)Saeidi, and van~der
  Weide]{saeidi2014wideband}
Saeidi,~C.; van~der Weide,~D. Wideband plasmonic focusing metasurfaces.
  \emph{Applied Physics Letters} \textbf{2014}, \emph{105}, 053107\relax
\mciteBstWouldAddEndPuncttrue
\mciteSetBstMidEndSepPunct{\mcitedefaultmidpunct}
{\mcitedefaultendpunct}{\mcitedefaultseppunct}\relax
\EndOfBibitem
\bibitem[Saeidi and van~der Weide(2015)Saeidi, and van~der
  Weide]{saeidi2015figure}
Saeidi,~C.; van~der Weide,~D. A figure of merit for focusing metasurfaces.
  \emph{Applied Physics Letters} \textbf{2015}, \emph{106}, 113110\relax
\mciteBstWouldAddEndPuncttrue
\mciteSetBstMidEndSepPunct{\mcitedefaultmidpunct}
{\mcitedefaultendpunct}{\mcitedefaultseppunct}\relax
\EndOfBibitem
\bibitem[Vakil and Engheta(2011)Vakil, and Engheta]{vakil2011transformation}
Vakil,~A.; Engheta,~N. Transformation optics using graphene. \emph{Science}
  \textbf{2011}, \emph{332}, 1291--1294\relax
\mciteBstWouldAddEndPuncttrue
\mciteSetBstMidEndSepPunct{\mcitedefaultmidpunct}
{\mcitedefaultendpunct}{\mcitedefaultseppunct}\relax
\EndOfBibitem
\bibitem[Vakil and Engheta(2012)Vakil, and Engheta]{vakil2012fourier}
Vakil,~A.; Engheta,~N. Fourier optics on graphene. \emph{Physical Review B}
  \textbf{2012}, \emph{85}, 075434\relax
\mciteBstWouldAddEndPuncttrue
\mciteSetBstMidEndSepPunct{\mcitedefaultmidpunct}
{\mcitedefaultendpunct}{\mcitedefaultseppunct}\relax
\EndOfBibitem
\bibitem[Tymchenko \latin{et~al.}(2013)Tymchenko, Nikitin, and
  Martin-Moreno]{tymchenko2013faraday}
Tymchenko,~M.; Nikitin,~A.~Y.; Martin-Moreno,~L. Faraday rotation due to
  excitation of magnetoplasmons in graphene microribbons. \emph{ACS nano}
  \textbf{2013}, \emph{7}, 9780--9787\relax
\mciteBstWouldAddEndPuncttrue
\mciteSetBstMidEndSepPunct{\mcitedefaultmidpunct}
{\mcitedefaultendpunct}{\mcitedefaultseppunct}\relax
\EndOfBibitem
\bibitem[Koppens \latin{et~al.}(2011)Koppens, Chang, and Garcia~de
  Abajo]{koppens2011graphene}
Koppens,~F.~H.; Chang,~D.~E.; Garcia~de Abajo,~F.~J. Graphene plasmonics: a
  platform for strong light--matter interactions. \emph{Nano letters}
  \textbf{2011}, \emph{11}, 3370--3377\relax
\mciteBstWouldAddEndPuncttrue
\mciteSetBstMidEndSepPunct{\mcitedefaultmidpunct}
{\mcitedefaultendpunct}{\mcitedefaultseppunct}\relax
\EndOfBibitem
\bibitem[Bao and Loh(2012)Bao, and Loh]{bao2012graphene}
Bao,~Q.; Loh,~K.~P. Graphene photonics, plasmonics, and broadband
  optoelectronic devices. \emph{ACS nano} \textbf{2012}, \emph{6},
  3677--3694\relax
\mciteBstWouldAddEndPuncttrue
\mciteSetBstMidEndSepPunct{\mcitedefaultmidpunct}
{\mcitedefaultendpunct}{\mcitedefaultseppunct}\relax
\EndOfBibitem
\bibitem[Lu \latin{et~al.}(2013)Lu, Zhu, Xu, Ni, Dong, and Cui]{lu2013flexible}
Lu,~W.~B.; Zhu,~W.; Xu,~H.~J.; Ni,~Z.~H.; Dong,~Z.~G.; Cui,~T.~J. Flexible
  transformation plasmonics using graphene. \emph{Optics express}
  \textbf{2013}, \emph{21}, 10475--10482\relax
\mciteBstWouldAddEndPuncttrue
\mciteSetBstMidEndSepPunct{\mcitedefaultmidpunct}
{\mcitedefaultendpunct}{\mcitedefaultseppunct}\relax
\EndOfBibitem
\bibitem[Chen and Al{\`u}(2011)Chen, and Al{\`u}]{chen2011atomically}
Chen,~P.-Y.; Al{\`u},~A. Atomically thin surface cloak using graphene
  monolayers. \emph{ACS nano} \textbf{2011}, \emph{5}, 5855--5863\relax
\mciteBstWouldAddEndPuncttrue
\mciteSetBstMidEndSepPunct{\mcitedefaultmidpunct}
{\mcitedefaultendpunct}{\mcitedefaultseppunct}\relax
\EndOfBibitem
\bibitem[Fallahi and Perruisseau-Carrier(2012)Fallahi, and
  Perruisseau-Carrier]{fallahi2012design}
Fallahi,~A.; Perruisseau-Carrier,~J. Design of tunable biperiodic graphene
  metasurfaces. \emph{Physical Review B} \textbf{2012}, \emph{86}, 195408\relax
\mciteBstWouldAddEndPuncttrue
\mciteSetBstMidEndSepPunct{\mcitedefaultmidpunct}
{\mcitedefaultendpunct}{\mcitedefaultseppunct}\relax
\EndOfBibitem
\bibitem[Wang \latin{et~al.}(2014)Wang, Liu, Lu, and Zeng]{wang2014graphene}
Wang,~G.; Liu,~X.; Lu,~H.; Zeng,~C. Graphene plasmonic lens for manipulating
  energy flow. \emph{Scientific reports} \textbf{2014}, \emph{4}\relax
\mciteBstWouldAddEndPuncttrue
\mciteSetBstMidEndSepPunct{\mcitedefaultmidpunct}
{\mcitedefaultendpunct}{\mcitedefaultseppunct}\relax
\EndOfBibitem
\bibitem[Pfeiffer and Grbic(2013)Pfeiffer, and Grbic]{pfeiffer2013cascaded}
Pfeiffer,~C.; Grbic,~A. Cascaded metasurfaces for complete phase and
  polarization control. \emph{Applied Physics Letters} \textbf{2013},
  \emph{102}, 231116\relax
\mciteBstWouldAddEndPuncttrue
\mciteSetBstMidEndSepPunct{\mcitedefaultmidpunct}
{\mcitedefaultendpunct}{\mcitedefaultseppunct}\relax
\EndOfBibitem
\bibitem[Pfeiffer \latin{et~al.}(2014)Pfeiffer, Emani, Shaltout, Boltasseva,
  Shalaev, and Grbic]{pfeiffer2014efficient}
Pfeiffer,~C.; Emani,~N.~K.; Shaltout,~A.~M.; Boltasseva,~A.; Shalaev,~V.~M.;
  Grbic,~A. Efficient light bending with isotropic metamaterial Huygens’
  surfaces. \emph{Nano letters} \textbf{2014}, \emph{14}, 2491--2497\relax
\mciteBstWouldAddEndPuncttrue
\mciteSetBstMidEndSepPunct{\mcitedefaultmidpunct}
{\mcitedefaultendpunct}{\mcitedefaultseppunct}\relax
\EndOfBibitem
\bibitem[Cheng and Mosallaei(2014)Cheng, and Mosallaei]{cheng2014optical}
Cheng,~J.; Mosallaei,~H. Optical metasurfaces for beam scanning in space.
  \emph{Optics letters} \textbf{2014}, \emph{39}, 2719--2722\relax
\mciteBstWouldAddEndPuncttrue
\mciteSetBstMidEndSepPunct{\mcitedefaultmidpunct}
{\mcitedefaultendpunct}{\mcitedefaultseppunct}\relax
\EndOfBibitem
\bibitem[Pors \latin{et~al.}(2013)Pors, Albrektsen, Radko, and
  Bozhevolnyi]{pors2013gap}
Pors,~A.; Albrektsen,~O.; Radko,~I.~P.; Bozhevolnyi,~S.~I. Gap plasmon-based
  metasurfaces for total control of reflected light. \emph{Scientific reports}
  \textbf{2013}, \emph{3}\relax
\mciteBstWouldAddEndPuncttrue
\mciteSetBstMidEndSepPunct{\mcitedefaultmidpunct}
{\mcitedefaultendpunct}{\mcitedefaultseppunct}\relax
\EndOfBibitem
\bibitem[Pors and Bozhevolnyi(2013)Pors, and Bozhevolnyi]{pors2013plasmonic}
Pors,~A.; Bozhevolnyi,~S.~I. Plasmonic metasurfaces for efficient phase control
  in reflection. \emph{Optics express} \textbf{2013}, \emph{21},
  27438--27451\relax
\mciteBstWouldAddEndPuncttrue
\mciteSetBstMidEndSepPunct{\mcitedefaultmidpunct}
{\mcitedefaultendpunct}{\mcitedefaultseppunct}\relax
\EndOfBibitem
\bibitem[Lau and Hum(2012)Lau, and Hum]{lau2012reconfigurable}
Lau,~J.~Y.; Hum,~S.~V. Reconfigurable transmitarray design approaches for
  beamforming applications. \emph{Antennas and Propagation, IEEE Transactions
  on} \textbf{2012}, \emph{60}, 5679--5689\relax
\mciteBstWouldAddEndPuncttrue
\mciteSetBstMidEndSepPunct{\mcitedefaultmidpunct}
{\mcitedefaultendpunct}{\mcitedefaultseppunct}\relax
\EndOfBibitem
\bibitem[Zentgraf \latin{et~al.}(2011)Zentgraf, Liu, Mikkelsen, Valentine, and
  Zhang]{zentgraf2011plasmonic}
Zentgraf,~T.; Liu,~Y.; Mikkelsen,~M.~H.; Valentine,~J.; Zhang,~X. Plasmonic
  luneburg and eaton lenses. \emph{Nature nanotechnology} \textbf{2011},
  \emph{6}, 151--155\relax
\mciteBstWouldAddEndPuncttrue
\mciteSetBstMidEndSepPunct{\mcitedefaultmidpunct}
{\mcitedefaultendpunct}{\mcitedefaultseppunct}\relax
\EndOfBibitem
\bibitem[Rejaei and Khavasi(2015)Rejaei, and Khavasi]{rejaei2015scattering}
Rejaei,~B.; Khavasi,~A. Scattering of surface plasmons on graphene by a
  discontinuity in surface conductivity. \emph{Journal of Optics}
  \textbf{2015}, \emph{17}, 075002\relax
\mciteBstWouldAddEndPuncttrue
\mciteSetBstMidEndSepPunct{\mcitedefaultmidpunct}
{\mcitedefaultendpunct}{\mcitedefaultseppunct}\relax
\EndOfBibitem
\bibitem[Orfanidis(2002)]{orfanidis2002electromagnetic}
Orfanidis,~S.~J. \emph{Electromagnetic waves and antennas}; Rutgers University
  New Brunswick, NJ, 2002\relax
\mciteBstWouldAddEndPuncttrue
\mciteSetBstMidEndSepPunct{\mcitedefaultmidpunct}
{\mcitedefaultendpunct}{\mcitedefaultseppunct}\relax
\EndOfBibitem
\bibitem[Hadad \latin{et~al.}(2014)Hadad, Davoyan, Engheta, and
  Steinberg]{hadad2014extreme}
Hadad,~Y.; Davoyan,~A.~R.; Engheta,~N.; Steinberg,~B.~Z. Extreme and quantized
  magneto-optics with graphene meta-atoms and metasurfaces. \emph{ACS
  Photonics} \textbf{2014}, \emph{1}, 1068--1073\relax
\mciteBstWouldAddEndPuncttrue
\mciteSetBstMidEndSepPunct{\mcitedefaultmidpunct}
{\mcitedefaultendpunct}{\mcitedefaultseppunct}\relax
\EndOfBibitem
\bibitem[Shen(2015)]{shen2015tunable}
Shen,~N.-H. Tunable terahertz meta-surface with graphene cut wires. \emph{ACS
  Photonics} \textbf{2015}, \relax
\mciteBstWouldAddEndPunctfalse
\mciteSetBstMidEndSepPunct{\mcitedefaultmidpunct}
{}{\mcitedefaultseppunct}\relax
\EndOfBibitem
\end{mcitethebibliography}

\providecommand{\latin}[1]{#1}
\providecommand*\mcitethebibliography{\thebibliography}
\csname @ifundefined\endcsname{endmcitethebibliography}
  {\let\endmcitethebibliography\endthebibliography}{}

\newpage



\end{document}